\documentclass[twocolumn,a4paper,9pt]{article}
\usepackage[utf8]{inputenc}
\usepackage[T1]{fontenc}
\usepackage{lmodern,textcomp}
\usepackage[english]{babel}
\usepackage[top=2.5cm,bottom=2.5cm,left=2.5cm,right=2.5cm,headsep=1cm]{geometry}
\usepackage{amsfonts,amsmath,amssymb}
\usepackage{color}
\usepackage{graphicx}
\usepackage{caption}
\usepackage{subfigure}
\usepackage{array}
\usepackage{multirow}
\usepackage{pict2e}
\usepackage{fancyhdr}
\title{Extreme events induced by collisions in a forced semiconductor laser}

\author{Pierre Walczak$^{1,*}$, Cristina Rimoldi$^{1}$, Francois Gustave$^{2}$, Lorenzo Columbo$^{3,4}$,\\Massimo Brambilla$^{4,5}$,Franco Prati$^{6,7}$,Giovanna Tissoni$^{1}$,Stéphane Barland$^{1}$}

\date{%
\small{$^{1}$Universit\'e Côte d'Azur, CNRS, Institut de Physique de Nice, Sophia Antipolis, France\\%
$^{2}$Univ. Lille, CNRS, UMR 8523 - PhLAM - Physique des Lasers Atomes et Molécules, Lille, France\\%
$^{3}$Politecnico di Torino, Dipartimento di Elettronica e Telecomunicazioni, Torino, Italy\\%
$^{4}$Istituto di Fotonica e Nanotecnologie del CNR, Bari, Italy\\%
$^{5}$Dipartimento di Fisica Interateneo, Universit\`a e Politecnico di Bari, Bari, Italy\\%
$^{6}$Dipartimento di Scienza e Alta Tecnologia, Universit\`a dell'Insubria, Como, Italy\\%
$^{7}$CNISM, Research Unit of Como, Como, Italy}\\%
$^{*}$Corresponding author: pierre.walczak@inphyni.cnrs.fr\\%
\today}

\begin{document}

\maketitle
\thispagestyle{fancy}

\textbf{We report on the experimental study of an optically driven multimode semiconductor laser with 1~m cavity length. We observed a spatiotemporal regime where real time measurements reveal very high intensity peaks in the laser field. Such a regime, which coexists with the locked state and with stable phase solitons, is characterized by the emergence of extreme events which produce a heavy tail statistics in the probability density function. We interpret the extreme events as collisions of spatiotemporal structures with opposite chirality. Numerical simulations of the semiconductor laser model, showing very similar dynamical behavior, substantiate our evidences and corroborate the description of such interactions as collisions between phase solitons and transient structures with different phase rotations.}
\vspace{1cm}

In the last decade, the understanding of extreme events as rogue waves arose considerable interest. These structures, consisting in a giant wave with a low probability of appearance, were first discovered in the context of oceanographic studies on the marine surface \cite{White:98}, and yet, the origin of rogue waves is still a matter of debate\cite{Ruban:10}. The common tool used to characterize the emergence of extreme events is the Probability Density Function (PDF). If the statistics deviates from the normal law, with a heavy tail PDF, then one speaks about rogue waves\cite{Onorato:13}.

Since the experiment of Solli \textit{et al.}\cite{Solli:07}, optics has become a useful workbench to study extreme events. Thanks to the analogy between optics and hydrodynamics, rogue waves have been studied in the one-dimensional nonlinear Schrödinger equation (1D NLSE) \cite{Agafontsev:2015,Chabchoub:15,Soto:16,Walczak:15,Suret:16}.  Although they may be hard to identify unequivocally \cite{Randoux:16,Soto:16}, different analytic solutions of the 1D NLSE such as solitons, rational solutions, solitons on finite background\cite{Akhmediev:09b,Dudley:14,Kibler:10,Kibler:12} or collisions between them \cite{frisquet:13} have been proposed as prototypes of rogue waves.

In dissipative systems, the phenomenon of optical rogue waves has been also studied, for example, in the supercontinuum generation \cite{Dudley:08}, in laser diodes with optical injection \cite{bonatto2011,Zamora:13} or optical feedback \cite{KarsaklianDalBosco:13,Reinoso:2013,Mercier:2015}, lasers with saturable absorber \cite{bonazzola2013,bonazzola2015,selmi2016,Coulibaly:17,rimoldi2017} and in resonant parametric oscillators \cite{Oppo:13}. In these contexts, the mechanism of extreme events formation is more complex to describe because of the presence of higher nonlinear terms, gain and losses. Different theories of rogue waves formation exist \cite{Kharif:03} where the presence of dissipation on the ocean surface is clearly present, and in fact some works concentrate on dissipative rogue waves \cite{kovalsky2011,Lacaplain:12}, although others focus more on the underlying spatiotemporal chaos \cite{Clerc:16,selmi2016,Coulibaly:17,rimoldi2017} within which rogue waves are observed. A means to gain control of such extreme events was proposed by exploiting instabilities and feedback in semiconductor lasers \cite{Akhmediev:16}. Dissipative optical rogue waves have been also studied in spatial domain (see \textit{e.g.} \cite{Arecchi:2011,Marsal:14,pierangeli2015spatial}). 

Recently, in the context of a forced semiconductor laser, we observed a great variety of regimes including frequency locked homogeneous state, turbulent regime and dissipative phase solitons \cite{Gustave:2015}. Since they emerge in a forced oscillatory medium, these solitons fundamentally consist of $2\pi$ phase rotations (whose direction sets a chirality) embedded in a phase locked background. In a propagative system with non instantaneous nonlinearity, only the counterclockwise direction is stable \cite{Gustave:16}. Upon collisions, these phase solitons can form complexes \cite{Gustave:17} with multiple chiral charge. In a recent paper \cite{Gibson:16}, the existence of extreme events has been associated to collisions of optical vortices with opposite charge, in two dimensional transverse systems showing vortex mediated turbulence, described by complex Ginzburg-Landau and Swift-Hohenberg models with external driving. 
Yet, the experimental observation of extreme events originated by collisions in an injected laser system remains a challenging task, and the present work bridges this gap in a temporal / propagative context.

In this letter, we experimentally show that collisions can create extreme events in a forced semiconductor laser. We also illustrate how these localized events with high amplitude modify the PDF of the power fluctuations of the wave. Furthermore, we performed numerical simulations based on a well-tested semiconductor laser model and provide a theoretical explanation for extreme events formation. 

\begin{figure}[htb]
\centering
\includegraphics[scale=0.65]{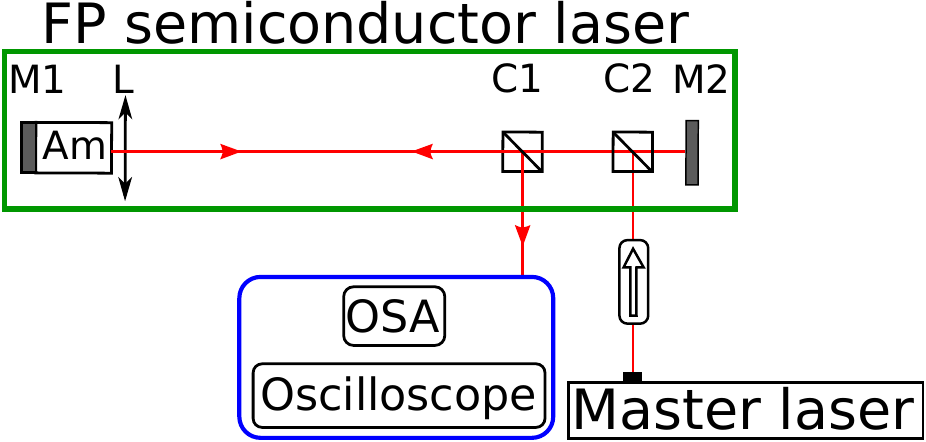}
\caption{Experimental setup. The Fabry Perot (FP) semiconductor laser is forced by a master laser. C1, C2: beam splitter (10\%); Am: active medium with antireflection coating on one side, and with highly reflective coating M1 on the other side; L: lens; M2: high reflectivity mirror (99\%); OSA: optical spectrum analyser. Cavity length is about 1~m.}
\label{fig1:setup}
\end{figure}

Fig.~\ref{fig1:setup} shows a schematic representation of the experimental setup implemented to investigate different dynamical regimes of a forced semiconductor laser. The slave laser is a semiconductor laser in a Fabry-Perot configuration where the corresponding power is called $P_{slave}$. The cavity is composed by an active medium which is antireflection coated only on one side. A lens is inserted to focus the output field on a high reflectivity mirror (99\%) which closes the one meter-long cavity. This cavity length has been chosen to allow a large number of longitudinal modes within the active medium's gain linewidth\cite{Gustave:2015} and a photon life time bigger than the carriers life time while still keeping a reasonably short round-trip time, enabling fully real-time measurements. Two 10\% beam splitters are inserted inside the Fabry-Perot resonator to provide both an input for the forcing field and an output for the detection of the emitted field. The master laser is provided by a grating tunable external cavity semiconductor laser. In order to prevent all reflections from the slave laser to the master laser, an optical isolator is inserted. The output is observed by using a high sensitive photodiode (Thorlabs PDA8GS) connected to a fast oscilloscope (Tektronix DPO71254C). Simultaneously, an optical spectrum analyser is used to measure the spectrum.

We use half-waveplate combined with a polarizing cube to control the power $P_{master}$ of the injected field inside the cavity. We obtain different dynamical regimes (stable phase solitons, turbulence or phase locking) depending on the detuning ($\Delta$) between the master and the slave laser. Here we focused on new regimes where an underlying irregular dynamics could couple with the propagation of self-organized structures. In particular we have selected a specific regime where we observe the emergence of a spatiotemporally localized event with a high amplitude.

\begin{figure}[htb]
\centering
\includegraphics[scale=0.38]{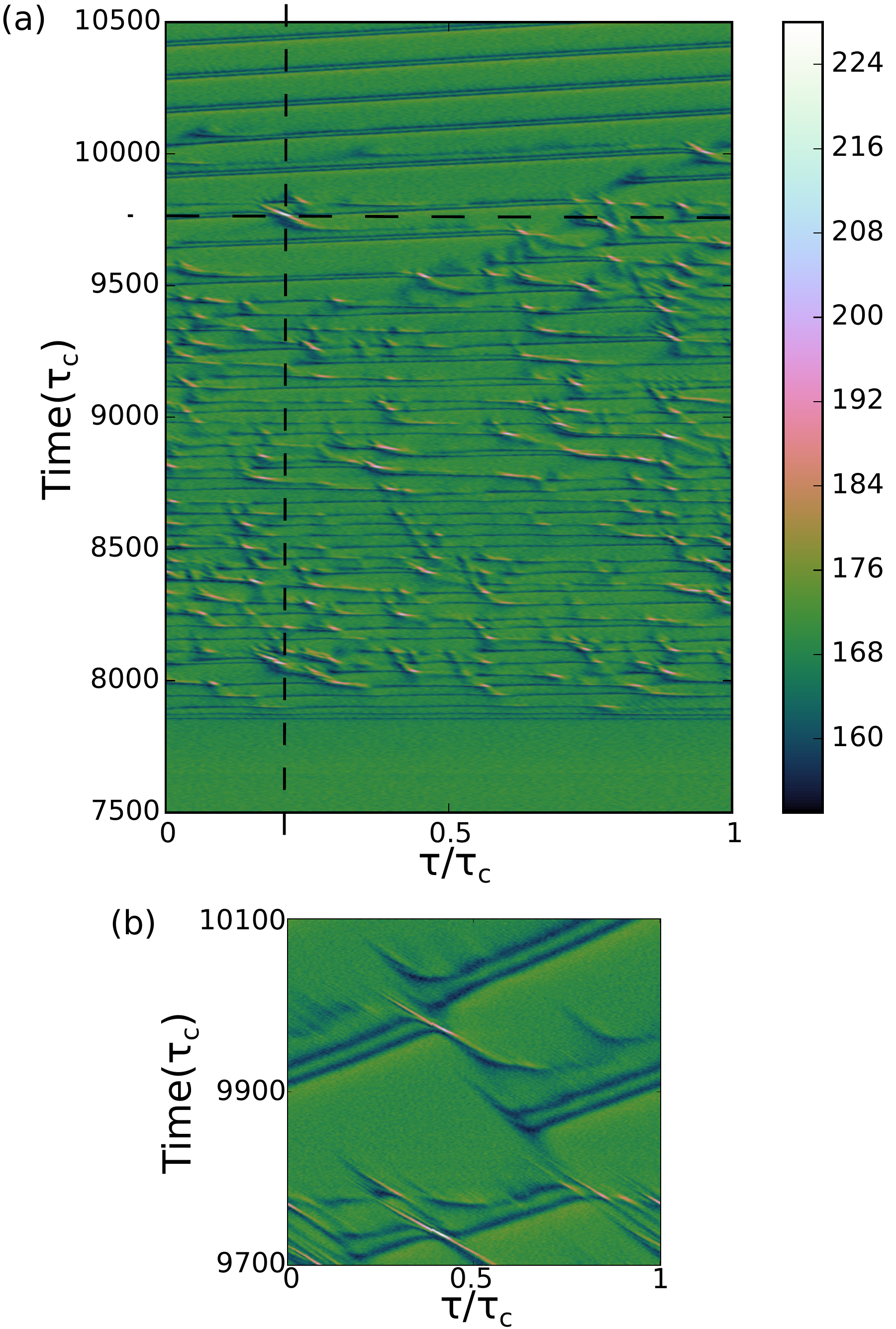}
\caption{(a) Space time diagram showing the different regimes observed in the experiment. Collisions between dissipative phase solitons and counter-propagating structures lead to the emergence of extreme events. Before 7500 the system is stably locked and after 10500 one soliton is stable. (b) Observation of collision in a slightly modified reference frame (6450~ps roundtrip time instead of 6470).  The laser threshold current is 18 mA, here its bias current is set at 1.6 times above the threshold. Experimental parameters: $P_{slave} = 700\mu W$ ; $P_{master} = 2.5 mW$ ; $\Delta=\lambda_{slave}-\lambda_{master} = -0.12 nm$. The actual power coupled into the cavity is at most a few \% of $P_{master}$.}
\label{fig2:diag_spatio_temp}
\end{figure}

\begin{figure}[htb]
\centering
\includegraphics[width=\linewidth]{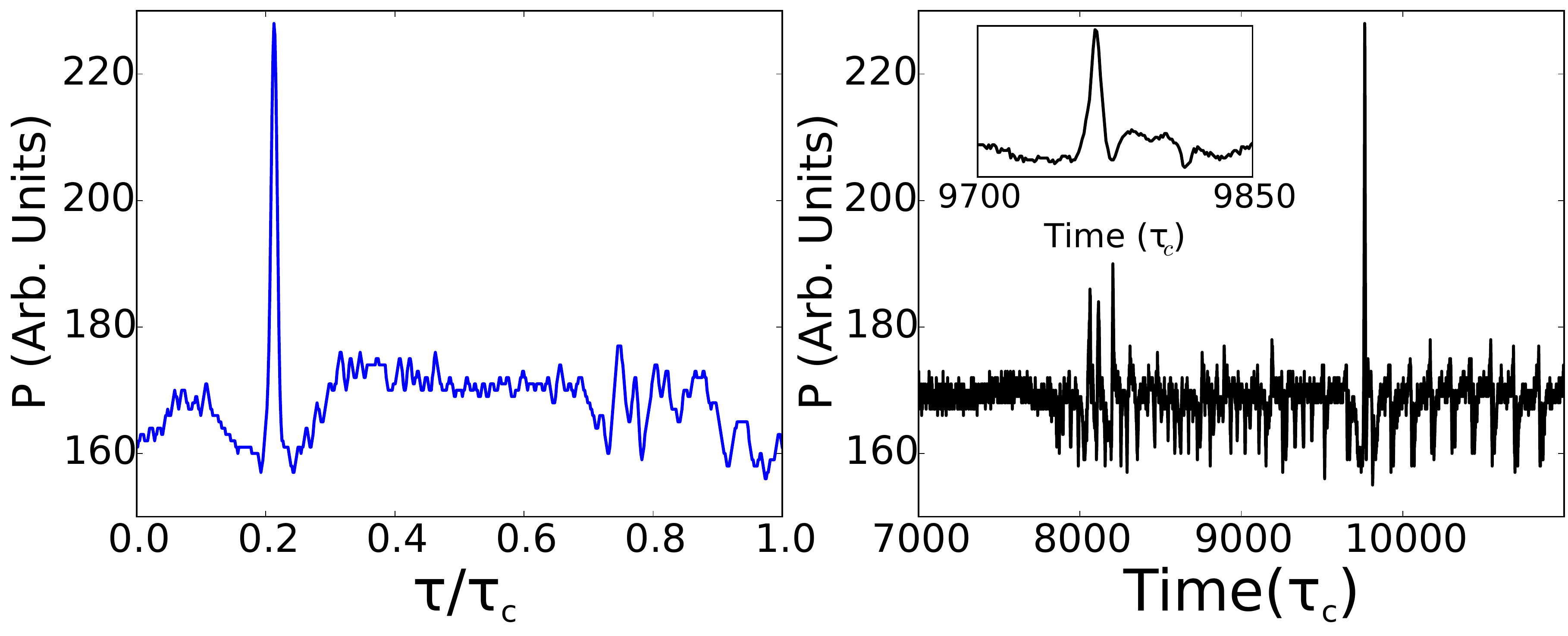}
\caption{Cross sections through the space time diagram localized around the event with the highest amplitude (corresponding to dashed black lines in Fig. \ref{fig2:diag_spatio_temp}). Left: Horizontal cut at the roundtrip 9764. Right: vertical cut at the time $\tau /\tau_{c}$ = 0.21.}
\label{fig3:coupe_fig17}
\end{figure}

Using the fast oscilloscope, we have recorded long time series of the intensity and constructed in real time a space-time diagram. To do that, we have folded the long time trace as a function of the roundtrip time of the cavity (6.45~ns). In this configuration, the horizontal axis is the roundtrip time, corresponding to the spatial dimension, and the vertical axis is the number of roundtrips. Fig.~\ref{fig2:diag_spatio_temp} (a) represents a spatiotemporal diagram in the regime described above. We can observe mainly three different behaviors. During the 7751 first roundtrips, the system is in synchronized state. Between roundtrips 7850 and 10060, the system exhibits a markedly different and complex dynamics. We can recognize the propagation of dissipative phase solitons, often with double charge. These coherent structures, which are spontaneously nucleated from the synchronized state, propagate and sometimes collide with different structures propagating in the opposite direction in this reference frame \footnote{Note that we measure only one direction of propagation of the field. Opposite propagation direction in the chosen reference frame does not mean opposite propagation direction of the field in the laboratory frame}. These collisions induce extreme events (see below for a statistical characterization) localized in time and space as we can see in Fig.~\ref{fig3:coupe_fig17}.

In Fig.~\ref{fig2:diag_spatio_temp} (b), we have chosen a slightly different comoving reference frame with respect to Fig.~\ref{fig2:diag_spatio_temp} (a) in order to optimize the representation of the collision. In this case, we can clearly observe that before the collision, the dissipative phase soliton with chiral charge two collides with a different transient structure. After this collision, one hump of the soliton complex seems to be annihilated and a new hump is regenerated about ten roundtrips later. This double structure propagates and eventually collides again with a counterpropagating structure, giving rise to a new extreme event. Finally, after this complex regime (from roundtrip number 10400 to the end of this record), a regular regime is attained, dissipative phase solitons with a chiral charge of two are stable and propagate without collisions.

\begin{figure}[htb]
\centering
\includegraphics[scale=0.35]{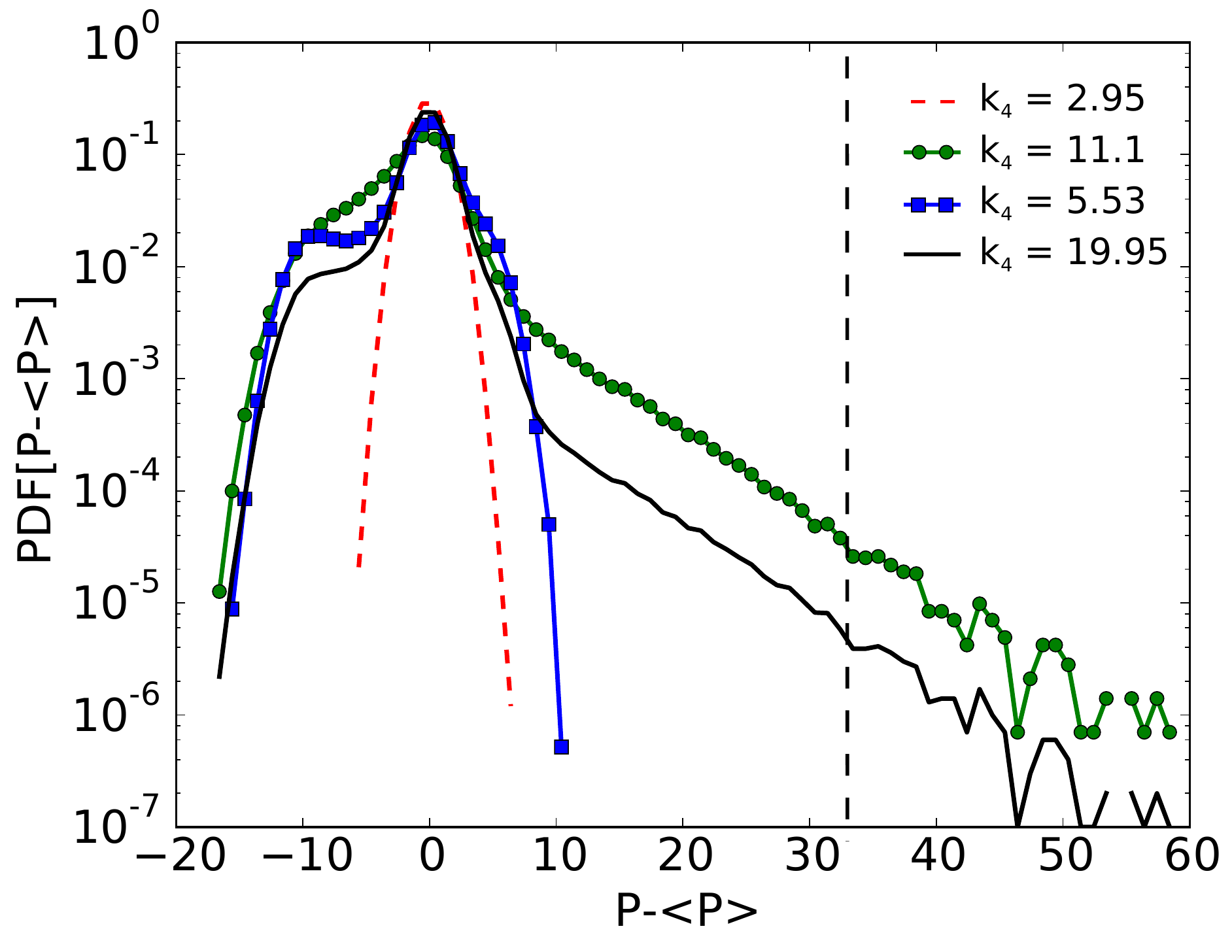}
\caption{Experiment: Probability Density Function (PDF) of the power fluctuations along the recording time trace. The PDF shows the deformation of the statistics with the emergence of localized events with high amplitude. Dashed red line : PDF from 0 to 7751 (locked state); Green line with circles: from 7850 to 10060 (collision regime); Blue line with squares:  from 10500 to 15500 (stable soliton propagation); Black line: PDF from 0 to 15500. Dashed black line: often used rogue wave threshold $<P>+8\sigma$ with $\sigma$ the standard deviation (calculated in the collision regime). Inset: kurtosis ($k_{4}$) for each PDF.}
\label{fig3:stat_fig17}
\end{figure}

The statistical impact of these events can be characterized by the computation of the PDF. Different approaches to PDF computation exist, depending on the flavour of the physics and the field of research\cite{Hammani:10,Onorato:13,Coulibaly:17}. In our system, we can have both complex dynamical states and completely coherent locked states. For this reason, we focus on the PDF of the power fluctuations for a long time trace as it is represented in Fig.~\ref{fig3:stat_fig17}.

The black solid line represents the PDF on all the roundtrips. The PDF is asymmetric and shows a clear heavy tail at the highest power values. The dashed red line, green line with circles and blue line with squares correspond respectively to the synchronized state, the turbulent regime and the soliton propagation regime described in Fig.~\ref{fig2:diag_spatio_temp} (a). In the synchronized state, where the emission is regular and continuous, the PDF results essentially from the physical and detection noise (dominant), which is stochastic. This explains why the PDF has a gaussian form. In the opposite case, in the turbulent regime, the emergence of extreme events modifies strongly the PDF which deviates from the gaussian statistics to exhibit a heavy tail statistics. Finally, the propagation of dissipative phase solitons plays also a significant role in the modification of the PDF. Due to their particular shape including a depression on the trailing edge \cite{Gustave:2015,Gustave:16}, the probability of the low power values is increased.

\normalsize{To simulate a semiconductor cavity with coherent injection we consider the following set of rate equations }

\scriptsize{
\begin{eqnarray*}
\frac{\partial E}{\partial z}+\frac{\partial E}{\partial t}-d\frac{\partial^2 E}{\partial z^2}&=&T\left[-(1+i\theta) E+y+(1-i\alpha)ED\right]\\
\frac{\partial D}{\partial t}&=&b\left[\mu-D\left(1+|E|^2\right)\right]
\end{eqnarray*}}
\normalsize{which is derived from the effective Maxwell-Bloch equations when the polarization of the medium is adiabatically eliminated. $E$ and $D$ represent, respectively, the electric field and the excess of carrier density with respect to transparency. $y$ is the amplitude of the injected field, $\theta$ is the frequency mismatch between the injected field and the closest empty cavity resonance, $T$ represents cavity losses. Finally $\alpha$ is the linewidth enhancement factor, $\mu$ is the pumping current and $b$ is the ratio of the roundtrip time to carrier lifetime. For further information on the derivation of the model and its description see Ref. \cite{Gustave:16,Gustave:17}.
The diffusion term $d$ has been introduced phenomenologically to take into account the finite gain linewidth of the semiconductor laser.}

We kept fixed the parameters $\alpha=3$, $T=0.3$, $b=10$, and $d=10^{-6}$, and set the pump current well above threshold $\mu=2$ ($\mu_{th}=1$). The detuning $\theta=-2.7$ and the injection amplitude $y=0.11$ were chosen so that the homogenous stationary solution is triple-valued, and the upper state is stable. Futhermore, the value of $y$, $\mu$ and $\theta$ are very close to the experiment. 

\begin{figure}[htb]
\centering
\includegraphics[width=\linewidth]{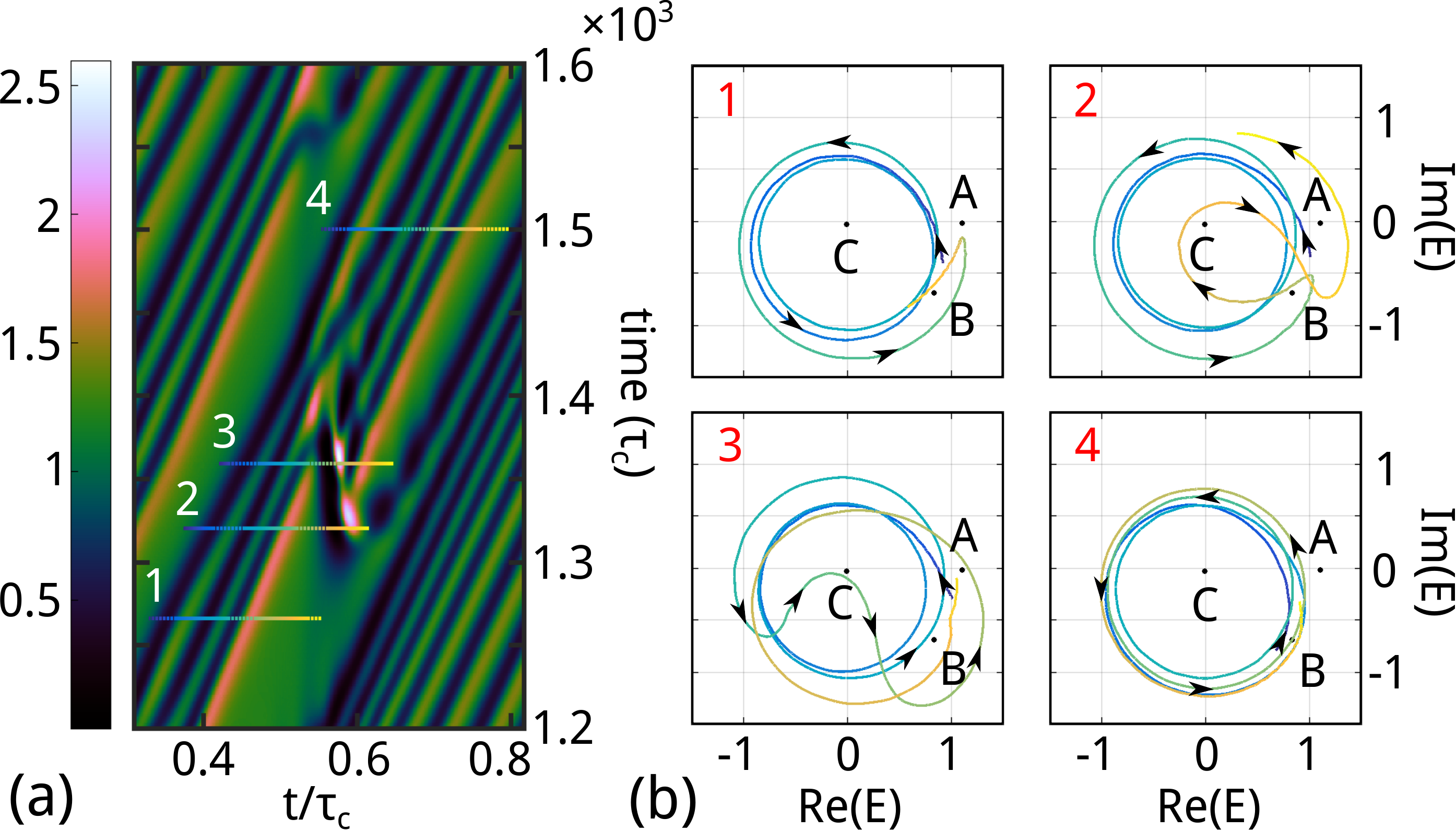}
\caption{Numerical simulation: zoom of the spatiotemporal diagram of the electric field intensity around a collision (a), and trajectories in the Argand plane for the selected roundtrip sections (b). Roundtrip section 3 coincides with the occurrence of a high intensity event and it is preceded by a clockwise rotation at the roundtrip section 2.}
\label{fig5:numerics}
\end{figure}

In Fig. \ref{fig5:numerics} (b) we show a zoom of the spatio-temporal diagram in a simulation where the initial condition was the homogeneous stationary solution with a superimposed phase kink of 4$\pi$ (but the same kind of regime can develop starting from noise). We can notice that the spatiotemporal dynamics present some clear similarities with the experimental data. In particular we can observe the presence of phase solitons and the emergence of some peculiar structures moving at a different velocity that eventually collide with the phase solitons, giving rise to an event of high intensity. 

In Fig. \ref{fig5:numerics}(a) we show the phase diagrams relative to the horizontal cuts highlighted in Fig. \ref{fig5:numerics}(b), time grows from red to yellow. The points $A$, $B$ and $C$ are the projections in the Argand plane of the the three fixed points \cite{Gustave:2015,Gustave:16}, respectively a node (stable for our choice of parameters), a saddle and an unstable focus. In the first frame, relative to roundtrip section 1, we can observe a phase soliton complex \cite{Gustave:17} of charge 3 that propagates inside the cavity: the trajectory of the system consists in three  counterclockwise phase rotations in the Argand plane, passing only once close to point $A$. At roundtrip section 2 a new object emerges at the right side of the soliton complex, together with a clockwise phase rotation (charge -1) in the Argand plane. The interaction between the two structures gives rise to a high intensity event at roundtrip section 3, where the clockwise rotation has already been lost. The interaction has also the effect of altering the phase soliton complex velocity inside the cavity, as can be noticed just after roundtrip section 3. At roundtrip section 4 the complex has regained its shape, with one additional charge, coming from the interaction.

The above evidences indicate that the appearance of these short-lived pulses with clockwise phase rotations that collide with stable phase solitons is the basic physical mechanism responsible for the extreme events observed in this system.

In conclusion, we have reported the experimental observation of extreme events due to the collision of localized structures in a forced semiconductor laser in a Fabry-Perot configuration. The appearance of these events localized in time and space changes strongly the PDF which exhibits a heavy tailed statistics. Numerical simulations based on a semiconductor laser model in a dynamical regime very close to the experiment have shown that strong intensity peaks follow the interaction between a phase soliton carrying a counterclockwise phase rotation and a low-intensity transient structure carrying a clockwise phase rotation. However, this transient object is of comparatively weak amplitude and its detection is beyond the capabilities of the phase measurement apparatus used in \cite{Gustave:2015,Gustave:16}.

\section*{Acknowledgments}
F.~G., G.~T., and S.~B. acknowledge funding from Agence Nationale de la Recherche through Grant No. ANR-12-JS04-0002-01. P.~W. and S.~B. acknowledge funding from Région Provence Alpes Côte d'Azur trough Grant No  15-1351.

\bibliographystyle{bibPi}
\bibliography{article}

\end{document}